\documentstyle[aas2pp4,psfig]{article}

\psfull

\long\def\Ignore#1{\relax}

\def\be{\begin{equation}}
\def\ee{\end{equation}}
\def\half{{\textstyle{1\over2}}}
\def\spose#1{\hbox to 0pt{#1\hss}} 
\def\ltsim{\mathrel{\spose{\lower 3pt\hbox{$\mathchar"218$}}
     \raise 2pt\hbox{$\mathchar"13C$}}}
\def\ltsim{\mathrel{\spose{\lower.5ex\hbox{$\mathchar"218$}}
     \raise.4ex\hbox{$\mathchar"13C$}}}
\def\gtsim{\mathrel{\spose{\lower.5ex \hbox{$\mathchar"218$}}
     \raise.4ex\hbox{$\mathchar"13E$}}}
\def\gtlt{\mathrel{\spose{\lower.5ex\hbox{$\mathchar"13E$}}
     \raise.5ex\hbox{$\mathchar"13C$}}}

\begin{document}

\title{Resonant Thickening of Disks by Small Satellite Galaxies}

\author{J.\ A.\ Sellwood}
\affil{Department of Physics and Astronomy, Rutgers, The State University of New Jersey, 136 Frelinghuysen Road, Piscataway, NJ 08854-8019}
\affil{Isaac Newton Institute, University of Cambridge, 20 Clarkson Road, Cambridge CB3 0EF, England}
\affil{sellwood@physics.rutgers.edu}\
\author{Robert W.\ Nelson}
\affil{Theoretical Astrophysics 130-33, California Institute of Technology, Pasadena CA 91125}
\affil{nelson@tapir.caltech.edu}
\author{and}
\author{Scott Tremaine}
\affil{Canadian Institute for Theoretical Astrophysics, University of Toronto, 60 St George Street, Toronto M5S 3H8, Canada}
\affil{Department of Astrophysical Sciences, Peyton Hall, Princeton University, Princeton, NJ 08544-1001} 
\affil{tremaine@astro.princeton.edu}

\begin{abstract}

We study the vertical heating and thickening of galaxy disks due to accretion
of small satellites.  Our simulations are restricted to axial symmetry,
which largely eliminates numerical evolution of the target galaxy but
requires the trajectory of the satellite to be along the symmetry axis of
the target.  We find that direct heating of disk stars by the satellite
is not important because the satellite's gravitational perturbation has
little power at frequencies resonant with the vertical stellar orbits.
The satellite does little damage to the disk until its decaying orbit
resonantly excites large-scale disk bending waves.  Bending waves can
damp through dynamical friction from the halo or internal wave-particle
resonances; we find that wave-particle resonances dominate the damping.
The principal vertical heating mechanism is therefore dissipation of
bending waves at resonances with stellar orbits in the disk.  Energy can
thus be deposited some distance from the point of impact of the satellite.
The net heating from a tightly bound satellite can be substantial, but
satellites that are tidally disrupted before they are able to excite
bending waves do not thicken the disk.

\keywords{galaxies: evolution --- galaxies: kinematics and dynamics ---
galaxies: internal structure --- galaxies: interactions  --- galaxies:
formation}

\end{abstract}

\section{Introduction}

Disk galaxies are observed to be cold and thin, with typical scale heights
only 10\% of their radial scale lengths.  The accretion of satellite
galaxies should strongly heat and thicken disks, so this observation
limits the satellite infall rate.  T\'oth \& Ostriker (1992; hereafter
TO) pointed out that thin galactic disks may therefore set important
cosmological constraints.

TO estimate the energy deposited in the disk during an accretion event in a
simplified manner.  They assume that the satellite galaxy spirals into the
parent galaxy on a near-circular orbit as it loses energy through dynamical
friction to both the dark matter halo and the disk.  They determine the
rate of energy loss to both components using Chandrasekhar's dynamical
friction formula (Binney \& Tremaine 1987, \S7.1), and deposit the energy
locally in the halo and disk, sharing the disk energy between vertical and
horizontal motions in a fixed ratio.  TO recognize that their treatment
omits all collective effects in the response, but argue that had they
``treated the problem as one of exciting modes in the disk$\ldots$[they]
would have found the same overall energy change in the disk.''

It is not obvious that this claim is correct; there are good reasons to
believe that in-plane heating will be increased while vertical heating
could be {\it reduced\/} by collective effects.  A swing-amplified spiral
response (see Binney \& Tremaine 1987 for a review) will always extract
energy from the potential well of the target galaxy, thereby adding to
the energy deposited into in-plane random motion.  On the other hand,
the following simple thought experiment suggests that vertical heating
could be small.  Let us imagine that the disk is very stiff so that the
eigenfrequencies of its collective bending modes are very high.  In this
case, the time-dependent perturbation of a passing satellite will not contain
power at frequencies which are resonant with any collective modes, and there
is little internal heating; the disk acts like a rigid plate and the only
effect of interaction with the satellite is to tilt or translate the disk.
We will develop these ideas further for more realistic disks in \S 2.

Many of TO's simplifying approximations are avoided in fully self-consistent
$N$-body simulations, which have been widely used to study the merger of
small satellites with a large disk galaxy (Quinn \& Goodman 1986; Pfenniger
1991; Quinn, Hernquist \& Fullagar 1993; Walker, Mihos \& Hernquist
1996; Athanassoula 1996; Huang \& Carlberg 1997).  These simulations
have mostly confirmed that disks are strongly heated by the accretion of
a $\sim10\%$ mass satellite and have elucidated other effects, such as
the stripping and tidal disruption of the satellite as it approaches the
target galaxy, and the tilting of the disk plane due to the gravitational
torque from the satellite.  Many of these calculations have begun with
models in only approximate equilibrium, and suffered from relaxation
and other numerical noise.  Numerical relaxation is reduced by particle
softening, but excessive softening impairs the ability of the disk to
support collective modes.  Of these simulations, that of Walker et al.\
was most successful at suppressing relaxation while maintaining a small
softening parameter, but their single simulation did not enable them to
reach a firm conclusion on many of the issues raised by TO.  Huang \&
Carlberg argue that TO overstate the disk heating associated with mergers;
they find that the disk absorbs some of the orbital angular momentum of
the satellite simply by tilting, which reduces the energy of vertical
oscillation available to thicken the disk.  These simulations have been
valuable, but a better appreciation of the physics of the heating process
is crucial if we are to understand how to generalize the simulation results.

Since it is the thinness of disks that is hardest to preserve, we focus
here on how vertical heating is affected by collective effects, and say
little about in-plane heating.  Moreover, vertical heating is a cleaner
problem than radial heating, since there is no radial redistribution of
the disk matter.  Vertical heating is largely unaffected by the energy
deposited into in-plane motion in the short run; it should be noted however,
that molecular clouds can gradually scatter horizontal motion into vertical
motion, thereby thickening the disk over timescales comparable to a Hubble
time (Carlberg 1987).

We first discuss heuristically, in \S 2, how energy might be deposited by
a satellite into vertical heat in the disk.  We test these ideas using
numerical simulations which are designed to avoid the complications
caused by numerical relaxation and rearrangement of angular momentum.
Our simulations utilize an axisymmetric grid and scarcely evolve
when unperturbed.  They avoid both internal relaxation and softening
while also being much faster than the direct $N$-body methods adopted in
previous studies.  We are therefore able to quantify the heating and to
explore more parameter space.  The assumption of axisymmetry restricts us
to mergers that occur along the symmetry axis, however.

There are several mitigating effects (tidal disruption of the satellite,
late star formation, infalling cold gas, etc.)\ that may reduce the
thickening caused by satellite mergers.  We do not address these issues
here, but focus on the detailed physical process of vertical heating by
a rigid massive satellite.

\section{A heuristic discussion of disk heating}

Chandrasekhar's dynamical friction formula assumes that the disk and
halo stars are free particles that travel on isotropically distributed,
straight-line orbits, and neglects the self-gravity of the response to
the perturbing satellite.  Any heating calculation based on that formula
thus ignores both the periodic orbital structure of the disk stars, and the
possibility of exciting large-scale collective modes in the form of bending
and density waves. We shall examine each of these effects in isolation,
even though they are closely related.

\subsection{Direct heating}

\begin{figure}[t]
\centerline{\psfig{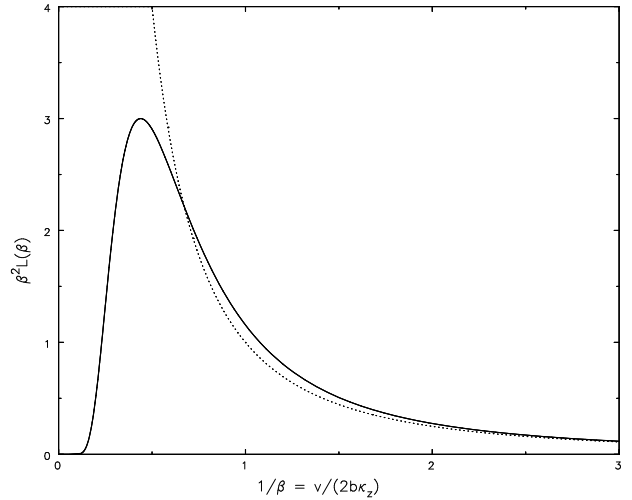}}

\caption{The dimensionless factor in the mean energy transferred to a
harmonically bound star (solid line) due to the tidal force of a passing
satellite (from equation 1).  For high velocities, the energy transfer
is the same as for a free particle (dotted line), but the transfer is
exponentially small at low velocities when the star responds adiabatically.}
\label{fig:hosc}

\end{figure}

We first consider an idealized problem that isolates the effects of
the periodic orbital structure of the disk stars. Stars near the disk
mid-plane oscillate vertically with a frequency $\kappa_z^2 = \half
\partial^2\Phi/\partial z^2$, where $\Phi({\bf x})$ is the gravitational
potential.  The tidal field from the passing satellite can be considered
to accelerate the star relative to the disk mid-plane.  For a satellite of
mass $M_s$ moving at relative speed $v$ and impact parameter $b$, one finds
the mean change in the vertical energy per unit mass of a disk star
during a single satellite passage with random orientation is (Spitzer 1958)
 \be
\Delta E_z = {h^2\over 3}\left ({GM_s\over b^3\kappa_z}\right )^2 \beta^2
L(\beta),
 \ee
where $h$ is the rms thickness of the disk and the parameter $\beta=2\kappa_z
b/v$ is of order the characteristic passage time of the satellite divided
by the orbital period of the star.  The dimensionless function $L(\beta)$
is unity for $\beta\to 0$ and exponentially small for $\beta\gg 1$.
Figure \ref{fig:hosc} shows that the net energy transfer peaks at $\beta
\sim 1$, when the forcing contains significant power near $\kappa_z$ and
thus is resonant with the stellar orbit.  At high velocities ($1/\beta
\gg 1$), the net energy transferred is nearly the same as for a free
particle, and thus varies as $\beta^2$ or $v^{-2}$.  At low velocities
($1/\beta \ll 1$), however, the star responds adiabatically and reversibly
so the energy transfer is exponentially small.  In other words, only stars
with impact parameters $b \ltsim v/\kappa_z$ gain energy from a satellite
passage.  TO's free-particle approximation effectively sets $\kappa_z\to0$
so that all stars gain energy.

The $h^2$ dependence in equation (1) indicates that the heating is tidal.
Direct heating vanishes as the disk becomes razor thin, both because $h
\rightarrow 0$ and also because $\kappa_z \rightarrow \infty$.

\subsection{Bending waves in razor-thin disks}

A self-gravitating disk supports collective modes, such as bending waves.
The simplest approximate analysis was given by Toomre (1966), who derived a
dispersion relation for bending waves in an idealized razor-thin membrane
with uniform surface density $\Sigma$ and horizontal velocity dispersion
$\sigma$; the self-gravity of the sheet resists vertical distortions in much
the same way as elasticity in conventional membranes.  Toomre's dispersion
relation reads
 \be
\omega^2=\nu_{\rm h}^2+2\pi G\Sigma |k|-k^2\sigma^2, \label{dispt}
 \ee
where $k$ is the horizontal wavenumber, and we have added a contribution
$\nu^2_{\rm h}=\half \partial^2 \Phi_h/\partial z^2$ from the vertical
frequency of free oscillations due to a halo potential $\Phi_{\rm h}$.
This dispersion relation says that the disk becomes unstable for sufficiently
large $k$, but in practice the instability is usually suppressed by nonzero
disk thickness.  Once excited, such waves propagate with group velocity
 \be
c_{\rm g}= {d\omega\over dk} = {\rm{sgn}}(k) {\pi G \Sigma\over \omega}\left
(1-{2k\over k_{\rm J}} \right ), \label{group}
 \ee
where $k_{\rm J} = 2\pi G\Sigma/\sigma^2$.

Equation (\ref{dispt}) suggests that long-wavelength modes have
eigenfrequencies $\omega \gtsim \nu_{\rm h}$.  The exception is the $m=1$
tilt mode (Sparke \& Casertano 1988), which would be neutral in a spherical
halo.  The gravitational field of the satellite excites bending waves if
and only if the Fourier transform of the external perturbation has power
at the mode frequencies.

What happens to energy deposited in Toomre's membrane in the form of
bending waves?  The wave cannot damp internally; it loses energy only
by dynamical friction with the halo (Dubinski \& Kuijken 1995, Nelson
\& Tremaine 1996).  Since direct heating is also negligible (\S 2.1),
a uniform membrane would not be heated at all by a satellite passage.

The modes of cold, disk-like membranes were investigated by Hunter \&
Toomre (1969).  Waves in a cold, axisymmetric disk may propagate all
the way to the disk edge.  Thus they found a spectrum of disk modes
containing both discrete and continuous eigenfrequencies, depending on
the sharpness of the disk edge.  In a membrane with non-zero horizontal
velocity dispersion, waves that would otherwise reach the disk edge may
encounter a turning point where $c_{\rm g}=0$ ($k= k_{\rm J}/2$ in equation
(\ref{group})) and can be trapped to form a discrete mode (Sellwood 1996).

What happens to the energy deposited in such systems by bending waves?
Once again, discrete modes damp only by dynamical friction with the halo,
and thus do not heat the disk. Continuous modes can also damp by nonlinear
effects that may become important if the mode propagates to the disk edge
(Hunter \& Toomre 1969); however, in this case only the edge of the disk
is heated.

In summary, a satellite passage does not heat razor-thin disks at all,
except possibly at their outer edge.

\subsection{Bending waves in realistic disks}

Razor-thin disks are unrealistic because all of the stars are forced to
move together in the vertical direction.  In real disks, stars oscillate
vertically at finite frequencies, allowing bending waves to damp through
wave-particle interactions (Landau damping).  The simplest model that
includes this effect is a planar distribution of stars which is infinite and
homogeneous in the $x$- and $y$-directions and held together by self-gravity
in the $z$ or vertical direction, with a velocity-dispersion tensor that
is independent of position.  The phase-space distribution function is
(Spitzer 1942, Camm 1950)
 \be
F(z,{\bf v})={\rho_0\over
(2\pi)^{3/2}\sigma^2\sigma_z}\exp\left(-{v_x^2+v_y^2\over 2\sigma^2}
-{E_z\over\sigma_z^2}\right), \label{spdf}
 \ee
where the vertical energy $E_z=\half v_z^2+\Phi(z)$, $\rho_0$ is the
mid-plane density, and $\sigma$ and $\sigma_z$ are constants equal to
the velocity dispersions in any horizontal direction and in $z$.  The rms
thickness for a Spitzer sheet $h \simeq 0.365 \sigma_z / (G\rho_0)^{1/2}$.
\Ignore{This is $1.8138z_0$, where $z_0 = \sigma_z / (8\pi G\rho_0)^{1/2}$}

A mathematical description of wave propagation in the Spitzer sheet
requires a solution of the linearized Boltzmann and Poisson equations for
the dispersion relation $\omega(k)$ (Toomre 1966, 1983, 1995; Araki 1985;
Weinberg 1991).  The result is not analytic and surprisingly more complicated
than equation (\ref{dispt}).  There are no real eigenfrequencies; all waves
are Landau damped.  Stars whose natural vertical oscillation frequency
$\kappa_z(E_z)$ resonates with the apparent (Doppler-shifted) frequency
$\omega-kv$ of the bending wave can absorb energy from the wave, converting
wave energy into random motion in the sheet.  In general, short-wavelength
modes $kh \gtsim 0.5$ damp in less than one wavelength (Toomre 1983, Weinberg
1991), while long-wavelength modes $kh \ll 1$ can propagate large distances.

The energy deposited in bending waves cannot be calculated accurately
without first knowing the shapes and frequencies of the bending eigenmodes
of the disk.  We expect, however, that the total energy deposited in a
large-scale bending mode with eigenfrequency $\omega$ during one passage
of a satellite is of order
 \be
\Delta E_w \sim M_d\left(GM_s\over va\right)^2{\cal L}(\beta^\prime),
 \ee
where $M_d$ is the disk mass, $a$ is the scale length of the disk,
$\beta^\prime = \omega a/v$, and ${\cal L}$ is exponentially small when
$\beta^\prime \gg 1$.  The major differences between direct heating (equation
1) and heating by excitation of bending waves are that (i) since the typical
modal frequency $\omega \ll \kappa_z$, as the satellite slows the stars
respond adiabatically and stop absorbing energy well before the modes;
(ii) since direct heating occurs through tidal forces, it is suppressed
by a factor $(h/a)^2$.

In disks where surface density and dispersions vary slowly with distance,
long-wavelength modes propagate until the wave enters a region where the
vertical frequency of the majority of stellar orbits is nearly resonant
with the wave frequency, $\omega \simeq \kappa_z(R)$.  Significant damping
occurs if the Doppler-shifted frequency is within one or two times the
`thermal width' of the resonance,
 \be
\left|{\hbox{Re}(\omega)-\kappa_z\over k\sigma}\right| \ltsim 1\hbox{
or }2.  \label{eq:sdef}
 \ee
Wave-particle interaction at vertical resonances is the primary mechanism
which converts the energy of a long-wavelength bending wave into internal
heat and thickens the disk.  Vertical resonances may occur at radii far
from the position where the wave is initially excited.

Exceptionally, the principal axis of the entire disk/halo can be re-aligned
by a satellite approaching off axis (Huang \& Carlberg 1997).  This new
equilibrium can be viewed as the trivial tilt mode with zero frequency,
which can be excited without causing significant heating.

\subsection{Summary}

These considerations suggest the following qualitative description
for the deposition of the satellite's orbital energy into vertical
motion in the disk:

{\it 1. An accreting satellite can deposit energy into the disk via
resonant excitation of either individual stellar orbits (``direct
heating'') or long-wavelength collective modes.}  Direct heating is
the closest analog to the energy transfer to free-particle orbits
envisaged by TO.  Collective modes include both (vertical) bending
waves and (horizontal) density waves, but we ignore the latter in the
present discussion because they do not thicken the disk directly.

{\it 2. At a given position in the disk, only the high-frequency
power $\omega \gtsim \kappa_z$ in the satellite force contributes to
direct heating.}  Little energy is deposited in high-speed passages
because the energy transfer $\propto v^{-2}$ in the impulsive limit.
In principle, direct heating is most important when the satellite
force is near resonance with the disk stars, but the excitation of
bending waves is more important in this regime.

{\it 3. Power at intermediate frequencies, $\nu_{\rm h} \ltsim \omega <
\kappa_{\rm z}$, can excite long-wavelength bending waves.}

{\it 4. For low-frequency perturbations, $\omega \ltsim \nu_{\rm h}$,
the disk responds adiabatically and reversibly.} The only exception
is that low-frequency perturbations can excite the tilt mode (Huang \&
Carlberg 1997).

{\it 5. Bending waves can damp by at least three mechanisms:  (a)
linear wave-particle resonances (Landau damping); (b) nonlinear
damping; (c) dynamical friction with the halo.}  Waves in real galaxies
can also be damped by other mechanisms, for example by viscosity or
shocks in the disk gas.  Gas dissipation is generally less effective
than other damping mechanisms because the mass in gas is much less than
the mass in stars except at the disk edge.  Only mechanisms (a) and
(b) heat the disk.

{\it 6. In contrast to the disk, the rate of direct heating of the halo
can generally be described by Chandrasekhar's formula.}  The reason is
that the halo contains a rich spectrum of orbital resonances with a
wide range of frequencies (Lin \& Tremaine 1983; Weinberg 1986).

\section{$N$-body model}

We use $N$-body simulations to assess the importance of collective
bending waves.  We simulate the interaction between a freely moving,
but internally rigid, satellite and an equilibrium disk-halo galaxy
model.  The satellite orbit lies along the symmetry axis of the
galaxy, so the galaxy response can be computed using a two-dimensional
axisymmetric grid code described by Sellwood (1996).  Since our rigid
satellites cannot be tidally stripped or disrupted, they settle intact
all the way to the galaxy center; the heating caused in this case is
clearly an upper limit.  In one separate experiment, we imagine the
satellite to dissolve before its orbit has completely decayed.

\begin{figure}[t]
\centerline{\psfig{figure=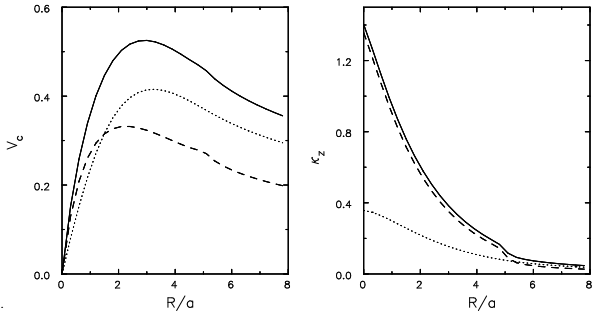,width=\hsize,clip=,angle=0}}

\caption{(left) The circular velocity and (right) the vertical
frequency for small oscillations, as functions of radius.  The dashed
and dotted curves show the separate contributions of the disk and
halo respectively and the full-drawn curves are for the total mass
distribution.  Thus, $\kappa_z$ is given by the full-drawn curve
on the right while $\nu_{\rm h}$ is given by the dotted curve.}
\label{fig:freq}

\end{figure}

The disk has an exponential surface-density distribution, contains
30\% of the total galaxy mass, $M$, and is truncated at $R=5a$,
where $a$ is the disk scale length.  We give the disk a constant
vertical thickness, generally $0.1a$ rms.  An exact equilibrium
disk model would require a distribution function (DF) that is a
function of three integrals, at least one of which is non-classical.
Since no such functions are available, we set Toomre's (1964) $Q=1.5$
to determine the radial velocity dispersion of the disk particles,
and estimate orbital velocities using the epicycle approximation
and Jeans equations.  We also set appropriate vertical velocities
to maintain the initial thickness.  Such a procedure becomes more
approximate as the ratio of the required radial velocity dispersion
to the circular velocity rises; this ratio is moderate in our case
and the initial model is acceptably close to equilibrium.

The remaining $0.7M$ of the galaxy is in a live and spherical
bulge/halo which, however, is not much more extended than the
disk.  The halo particles are selected from a DF; the procedure
for constructing an equilibrium DF in the potential of both the
disk and halo will be described in a future paper (Debattista \&
Sellwood, in preparation).  We have opted for a simple isotropic DF
of polytropic form
 \be
f(E) = \cases{ (\Phi_* - E)^{n-3/2} & $ E \leq \Phi_*$ \cr 0, &
otherwise \cr}
 \ee
where $n$ is the polytropic index and $\Phi_*$ is the potential at
which the halo density is desired to drop to zero.  For a polytrope,
the density is proportional to the $n$-th power of the relative
potential (Binney \& Tremaine 1987, \S 4.4); we choose $n = 2$.
We choose particles from this DF using the procedure described in
Sellwood \& Valluri (1997).

Overall, the model is close to an exact detailed equilibrium; the worst
blemish is a mild imbalance at the outer edge of the disk causing it
to spread be\-yond the initial truncation radius by a few percent.
Unperturbed models have been verified to be stable and, after some
slight adjustment of the disk edge, do not evolve (as will be shown
in Figure {\ref{fig:vaporized}).

\begin{table}[t]
\setlength{\baselineskip}{12pt}
\def\fm{\footnotemark}
\setlength{\baselineskip}{12pt}
\centering
\begin{tabular}{lc}
   \multicolumn{2}{c}{Table 1. 
    Galaxy Model Parameters} \\ \hline \hline
   \multicolumn{1}{c}{Quantity} & 
   \multicolumn{1}{c}{Value}\\ \hline 
   \multicolumn{2}{c}{Disk} \\ \hline 
   Model \dotfill  & exponential \\
   Mass $M_d$ \dotfill &  $0.3M$  \\
   RMS vertical thickness $h$ \dotfill & $0.1a$ \\
   Toomre $Q$ \dotfill   & $1.5$  \\ 
   Truncation radius \dotfill  & $5a$  \\ \hline
   \multicolumn{2}{c}{Halo} \\ \hline 
   Model \dotfill  & $n=2$ polytrope \\
   Mass $M_h$ \dotfill &  $0.7M$  \\
   Truncation radius \dotfill  & $5.62a$  \\ \hline
   \multicolumn{2}{c}{Satellite} \\ \hline 
   Model \dotfill  & rigid Plummer sphere \\
   Mass $M_s$ \dotfill & $0.05 M$  \\ 
   Core radius $r_c$ \dotfill  & $0.125 a$  \\ \hline \hline
   \multicolumn{2}{c}{Numerical Parameters} \\ \hline \hline
   Number of disk particles \dotfill  & $10^5$ \\
   Radial grid spacing \dotfill &  $0.1a$  \\
   Vertical grid spacing \dotfill & $0.01a$ \\
   Time step \dotfill   & $0.02(a^3/GM)^{1/2}$  \\ \hline
\label{tab:one}
\end{tabular}
\end{table}

Figure \ref{fig:freq} shows the rotation curve as well as the radial run of
$\kappa_z$ and $\nu_h$, the frequency of small vertical oscillations in the
mid-plane in respectively the total potential and that of the halo only.
Our model is somewhat unrealistic for a galaxy since we were forced
to choose $\Phi_*$ to be the potential at the grid edge, which severely
truncates the halo and leads to a falling rotation curve in the outer disk.
We justify using it on the grounds that we wish to study physical principles
rather than to make detailed comparisons with real galaxies.  We also note
that the shape of the rotation curve is less important for axisymmetric
disturbances of the kind studied here than it is for non-axisymmetric ones.

In most experiments, the satellite is a rigid Plummer sphere having a
mass of typically $0.05M$ and a core radius $r_c=0.125a$.  Its motion is
computed from the negative of the vector sum of the forces it exerts on
all the particles, which are calculated directly and not through the grid.
We generally assume that the satellite has fallen from rest at infinity
and begin the simulation when it is at a distance of typically $10a$; up
to this moment the galaxy is assumed to have responded to the satellite's
approach as a rigid body.  In order to prevent large parts of the galaxy from
leaving the grid, we shift the grid position at every step by the amount
required to keep the center-of-mass of the galaxy centered on the grid.
This strategy successfully preserves most particles within the grid; no
more than 4\% of particles, all from the halo, leave the grid by the end
of a freely falling satellite run.

The grid spacing is uniform in both the radial and vertical directions;
there are 111 radial nodes spaced at $\delta R=0.1a$ while the 1125 vertical
nodes are spaced with $\delta z= 0.01a$.  The vertical structure of the
disk is well resolved.  We employ $100\,000$ equal-mass particles and
use a time-centered leap-frog integration scheme with $\delta t = 0.02
(a^3/GM)^{1/2}$.  Energy is conserved to better than 0.1\%.  The model
and numerical parameters are summarized in Table 1.

We have verified that our results are independent of the specific choices
for the numerical parameters.  We have also obtained similar results
using other methods to determine the gravitational field.  For checks
with fully three-dimensional methods, we substantially suppressed angular
momentum transport in the disk by reversing the sense of rotation of
half the particles.  The disks in these lower-resolution codes thickened
a little more, which is to be expected because the vertical cohesion of
the disk is diminished when forces are not so well resolved.

All quantities below, and in the Figures, are expressed in units such
that $G=M=a=1$.

\section{Disk heating by freely falling satellites}

\subsection{Numerical results}

In our first experiment (``standard run'') a satellite of mass $0.05M$
approaches the target galaxy down the symmetry axis on a parabolic orbit.
The $z$-distance of the satellite from the center of mass of the galaxy is
shown in figure \ref{fig:free}a.  Figure \ref{fig:free}b shows that the
satellite loses energy through dynamical friction in a series of steps,
each occurring as it passes through the densest part of the galaxy.
Figure \ref{fig:free}c shows the time-dependent vertical force as seen
by a star at the center of mass of the galaxy.  The maximum force during
an encounter is $2GM_s/(3^{3/2}r^2_c)= 1.23$ in our units.  The initial
passages produce a sequence of well-separated impulsive spikes; at late
times, as the satellite settles into the disk plane, the spikes become
broader and more closely spaced.

\begin{figure}[t]
\centerline{\psfig{figure=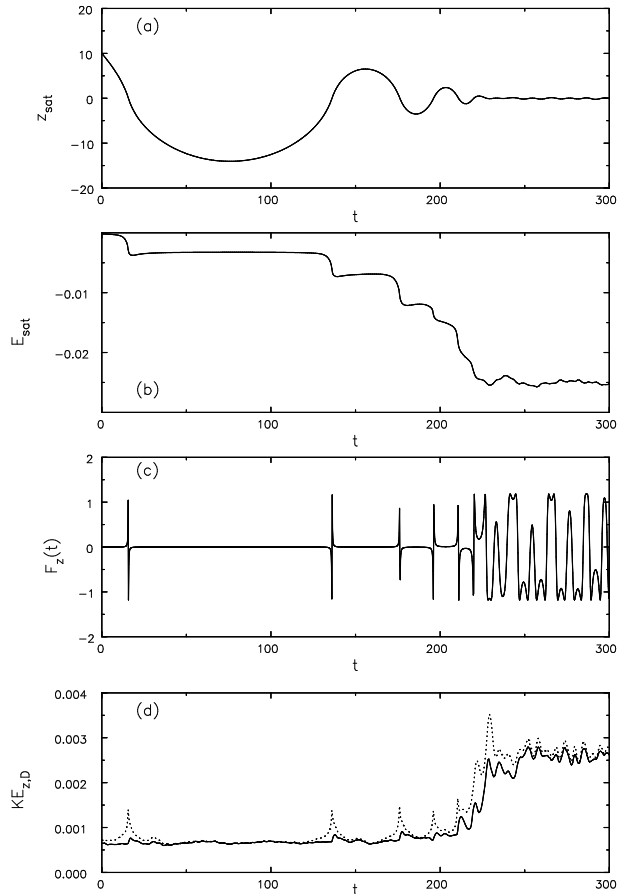,width=\hsize,clip=,angle=0}}

\caption{The orbital decay in our standard run, in which a 5\% satellite
falls down the symmetry axis of the target galaxy.  (a) The distance between
the satellite center and the center of mass of the target galaxy.  (b) The
kinetic plus potential energy of the satellite, measured in the barycentric
frame of the galaxy plus satellite.  (c) The vertical acceleration due
to the satellite felt by a star at the center of mass of the galaxy.
(d) The kinetic energy of vertical motion of the disk particles (dotted)
and the kinetic energy of random vertical motion (solid curve).}
\label{fig:free}
\end{figure}

The vertical heating is shown in Figure \ref{fig:free}d.  The dotted curve
shows the kinetic energy in vertical motion, defined in the frame of the
barycenter of the galaxy and satellite.  The solid curve shows the kinetic
energy of motion relative to the mean velocity of the nearby disk stars.
Almost all of the satellite energy is lost to the halo at first and only
after the fourth disk passage at $t \simeq 200$ does the disk begin to
heat significantly.

The first passage of the satellite through the disk produces a mild
example of a ring formed through the mechanism discussed by Lynds \& Toomre
(1976).  The ring is quite indistinct, however, and the increase in the
horizontal random kinetic energy is just a few percent as a result of this
first passage.  There is little secular increase of the horizontal random
motion within the disk until after $t \simeq 200$.  In the late stages of
the merger, however, the horizontal kinetic energy of random motion doubles.

The substantial vertical heating at late times is associated with violent
convulsions of the disk, as shown in Figure \ref{fig:violent}, after
which time the rms disk thickness is some four times its original value.
Note that only about 8\% of the energy lost by the satellite ends up
as increased kinetic energy of vertical motion of the disk particles;
the damage caused by this small fraction of the total available energy
underscores the extreme fragility of disks.

\begin{figure}[t]
\centerline{\psfig{figure=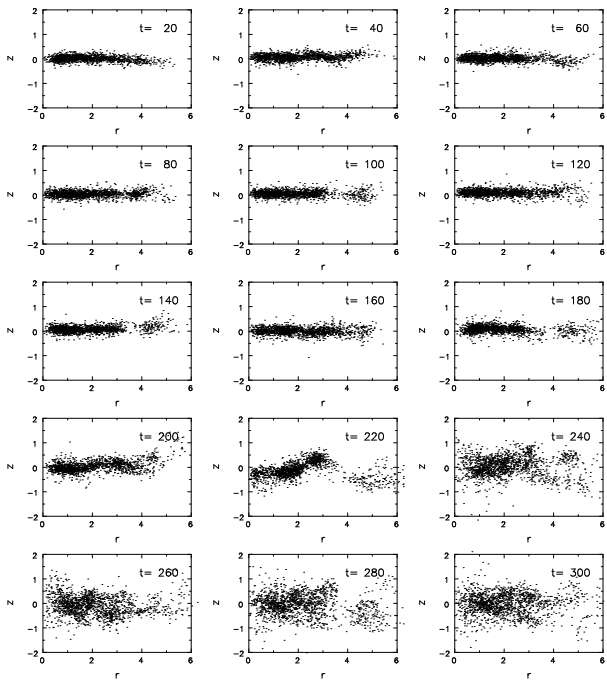,width=\hsize,clip=,angle=0}}

\caption{Meridional projections of 1500 representative disk particles at
equally spaced times.  The satellite has passed through the disk three times
by $t=180$ yet the disk has suffered little damage.  After $t=200$, however,
the disk convulses dramatically and quickly thickens.}
\label{fig:violent}
\end{figure}

\begin{figure}[t]

\centerline{\psfig{figure=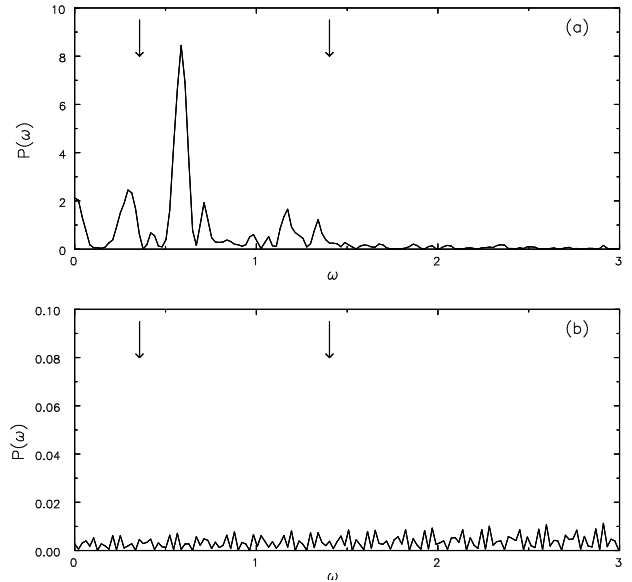,width=\hsize,clip=,angle=0}}

\caption{(a) The power spectrum of the vertical force shown in Figure
\protect{\ref{fig:free}c}; (b) the power spectrum if the force in
\protect{\ref{fig:free}c} is multiplied by a window which is unity for $t <
150$ and decays smoothly to zero (as $\cos^2$) between $t=150$ and $t=200$.
The vertical arrows at $\omega \simeq 0.35$ and $1.3$ denote the central
values of $\nu_h$ and $\kappa_z$ respectively.}
\label{fig:fourier}
\end{figure}

\begin{figure}[t]

\centerline{\psfig{figure=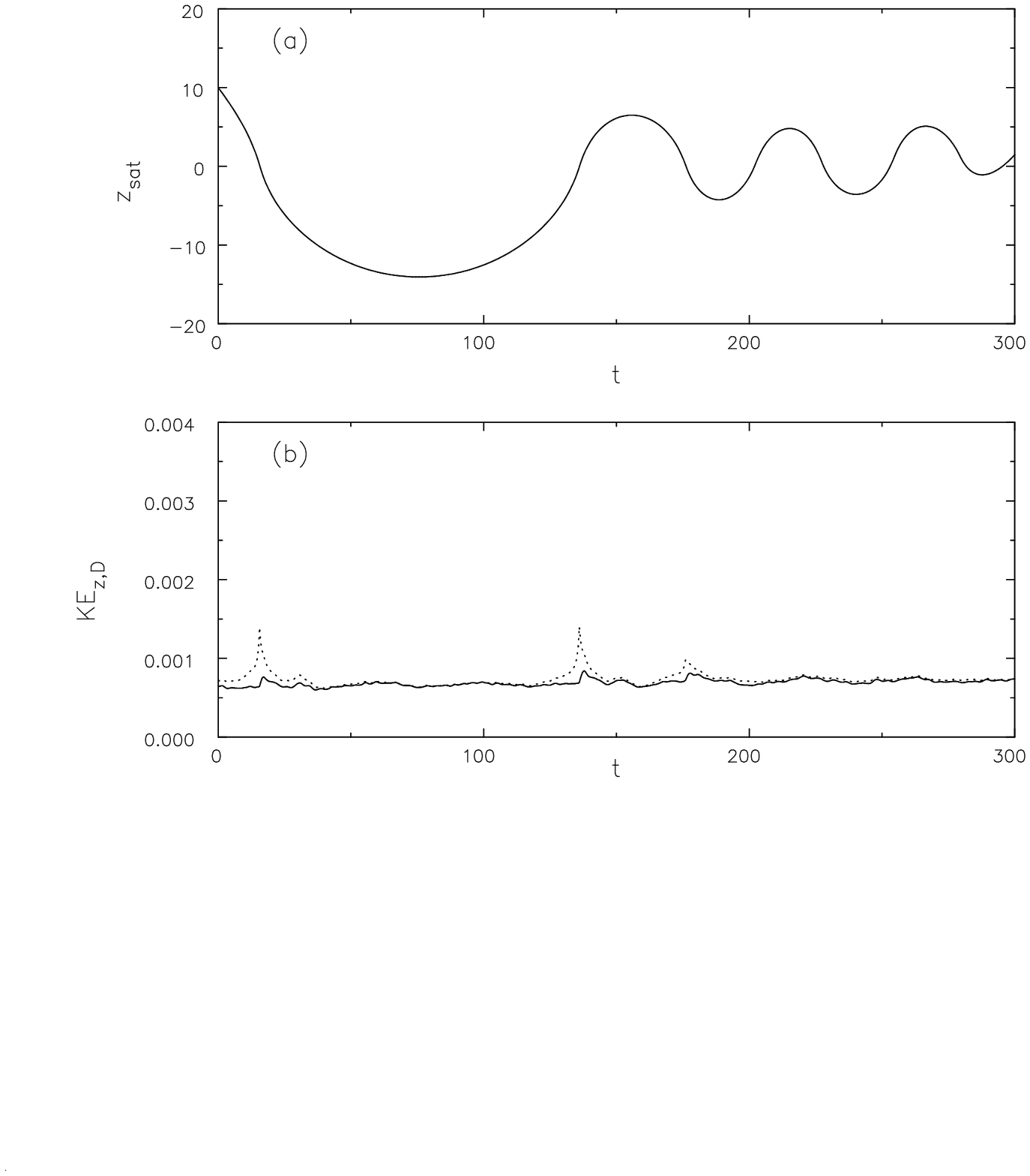,width=\hsize,clip=,angle=0}}

\caption{As for Figure \protect{\ref{fig:free}a} and
\protect{\ref{fig:free}d}, but for an experiment in which the satellite was
``vaporized'' over the time interval $150 \leq t \leq 200$.  The vertical
heating of the disk is barely detectable in this case.  The residual
particle experiences no friction after $t=200$ because of its negligible
mass.  The lower panel also demonstrates that an unperturbed disk does
not thicken over the course of an experiment.}
\label{fig:vaporized}
\end{figure}

We have run other experiments starting with both thicker and thinner disks,
as well as experiments with other satellite masses.  Results from a run with
an initially thinner disk were very similar; the orbit decayed at almost
exactly the same rate and the final thickness was only marginally less.
A thicker disk was less robust, however; the satellite orbit decayed slightly
more quickly and the final disk thickness was substantially greater than
in the end state of the standard run.  Experiments with other satellite
masses did not reveal any surprises.

\subsection{Interpretation}

We are now in a position to interpret these results in terms of the physical
processes described in \S 2.  The satellite velocity during the first disk
passage is $v \sim 1$, so a star at the center feels perturbations with
frequencies up to $\omega \sim v/r_c=8$.  This exceeds $\kappa_z=1.3$
at $R=0$ (Figure \ref{fig:freq}), so the heating is direct, but little
heating occurs during these fast passages because the net energy transfer
$\Delta E \propto v^{-2}$.  At late stages, when the satellite has nearly
settled into the disk, it oscillates in $z$ with a frequency $\sim 0.6$
which is less than $\kappa_z(0)$; therefore there is little or no direct
heating, but the satellite can excite bending waves.

The energy input into bending waves is indicated by the difference between
the dotted and solid curves in Figure \ref{fig:free}d.  Whenever the total
energy in vertical motion is larger than the energy in random vertical
motion, the difference is approximately the energy in bending waves.
Increases in random vertical motion generally lag the total vertical energy,
perhaps by $\sim 10$ time units, suggesting that the waves damp sometime
after they are excited.  This delay is consistent with the propagation time
across the disk as given by a typical group velocity (equation \ref{group}).

Figure \ref{fig:fourier}a shows the power spectrum of the perturbing
force experienced by a particle at the disk center, evaluated from the
data shown Figure \ref{fig:free}c.  The strong peak at $\omega \simeq 0.6$
arises mainly from oscillations of the satellite at late times $t > 240$.
There is little power at $\omega \gtsim \kappa_z(0) \simeq 1.3$, which
again shows why direct heating is inefficient.  There is significant power
at intermediate frequencies $\omega \gtsim \nu_h \simeq 0.3$, which excites
bending waves.  The excitation of bending waves mostly occurs at late times
when the satellite orbital frequency resonates with the long-wavelength disk
modes.  To illustrate this, in Figure \ref{fig:fourier}b we have plotted the
power spectrum that would arise if the force in Figure \ref{fig:free}c were
multiplied by a window that is unity at early times and decays smoothly
to $0$ between $t=150-200$; the strong low-frequency power is removed.
Note that the scales in the two panels differ by a factor of 100.

To confirm that most of the heating occurs at late times we have conducted an
experiment identical to the standard run up until time 150; we then gradually
``vaporized'' the satellite (with a $\cos^2$ time dependence) from its
initial 5\% of the galaxy mass down to that of all the other particles.
The results are shown in Figure \ref{fig:vaporized}.  Even though the
satellite passed through the disk twice when it had its full mass and a
third time with about half its initial mass, the increase in the vertical
random kinetic energy of the disk particles is barely detectable. This
result is consistent with the conclusion of Huang \& Carlberg (1997) that
a satellite does not heat the disk at all if it is tidally disrupted before
it reaches the vicinity of the disk.

\section{Additional numerical experiments}

\subsection{Off-center encounters}

We are unable to study off-center impacts properly with an axisymmetric
$N-$body code.  Nevertheless, we may consider the interaction of a massive
ring-like satellite with the galaxy -- an admittedly artificial arrangement,
but one which is still instructive.

The ``satellite'' in these experiments is a massive ring having 5\% of
the galaxy mass.  The ring initially has no radial or vertical velocity
($v_z=v_R=0$), but $v_\phi$ is set so that the ring has sufficient angular
momentum to balance the gravitational attraction towards the symmetry axis.
The ring mass is added to the calculation grid so that the mutual forces
between the satellite and the particles can be calculated using the grid.
Because there are no tangential forces in our experiments, the satellite
cannot lose angular momentum, but it can give up energy from its vertical
motion to the lighter particles of the disk and halo.

\begin{figure}[t]

\centerline{\psfig{figure=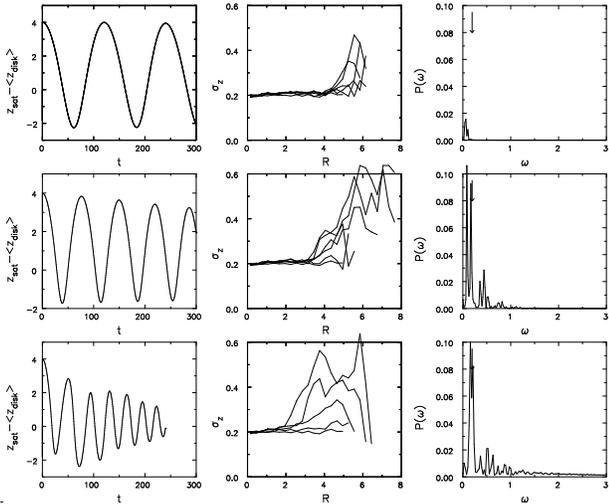,width=\hsize,clip=,angle=270}}

\caption{Three experiments with ring-like satellites showing the vertical
motion of the satellite on the left.  The different curves in each center
panel show the rms disk thickness as a function of radius at intervals of
60 time units.  The power spectrum of the force experienced by a particle
at $(R,z)=(4,0)$ is shown on the right.  The satellite started at $z=4$ in
each case and at $R=8$ (top row), $R=6$ (middle row) and $R=4$ (bottom row).}
\label{fig:rings}
\end{figure}

We initially performed these ring experiments using the same disk and halo
described in \S 3.  In this case we found the satellite did little heating
at all.  For this disk, with thickness $0.1a$, the vertical forcing has
little power above the stellar vertical frequencies, $\kappa_z$ (see figures
\ref{fig:freq} and \ref{fig:fourier}).  Consequently, we use a flabbier disk
with double the thickness for these experiments.  Figure \ref{fig:rings}
shows results from three simulations in which the perturber was started at
$(R,z) = (8,4)$, $(6,4)$ and $(4,4)$. Recall that the disk edge is at $R=5$
and the halo mass density is very low beyond this radius.

The satellite started outside the galaxy with $R=8$ oscillates about the
mid-plane with hardly any decay and little energy is transferred to the
disk.  The satellite started at $R=6$ decays slowly while the disk begins
to thicken near the outer edge only.  Finally, the orbit of the satellite
started at $R=4$, which passes through the disk, decays more quickly and
disk thickening gradually extends inwards as the orbital period drops.
In the top panel, the satellite is well outside the truncation radii of
both the disk and the halo and it moves so slowly that the particles can
respond adiabatically.  As the forcing frequency rises, the perturbation
approaches resonance with some particles and the satellite loses energy.

Figure \ref{fig:rings} shows that disk particles that are close to resonance
absorb energy from the perturber, but it should be noted that most of the
energy lost by the satellite is absorbed by the halo particles.

\subsection{Forced response of a disk galaxy}
\label{sec:forced}

In this subsection we consider the response of a disk galaxy to a periodic
forced oscillation -- in this case, a satellite is driven up and down
the rotation axis at a fixed frequency. Although highly unrealistic,
this experiment illustrates the importance that vertical resonances play
in heating of the disk.

The mass of the perturber in these experiments is $0.01M$ and it is forced
to move as
 \be
z(t) = 0.05a\sin\left( {2\pi t \over \tau} \right),
 \ee
with $\tau$ being the oscillation period.  The perturbing mass starts in
the mid-plane at $t=0$ and the negative of its initial momentum is shared
equally among all the particles of the galaxy.  Figure \ref{fig:forced}
shows the result from an experiment with $\tau = 20$ in our time units.
The gravitational coupling between the forced perturber and the surrounding
disk is strong enough that the disk center is locked to it; the perturber
therefore launches an axisymmetric bending wave that travels radially
outwards through the disk.  The upward slope of the crests in the left-hand
panel shows that the driven bending wave propagates outwards, as expected for
WKB bending waves described by equations (\ref{dispt}) and (\ref{group}).
The amplitude increases with radius for some distance, as conservation
of wave action demands, but then drops again just after the point where
localized thickening of the disk occurs, as shown in the right-hand panel.

\begin{figure}[t]

\centerline{\psfig{figure=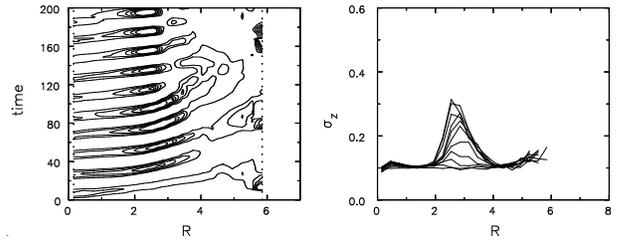,width=\hsize,clip=,angle=0}}

\caption{An experiment with a forced perturbation at period $\tau=20$
time units.  On the left we contour the mean $z$ displacement of the disk
particles as functions of radius and time; the contours show only positive
displacements.  The different curves in each right hand panel show the
rms disk thickness as a function of radius at intervals of 20 time units.}
\label{fig:forced}
\end{figure}

As described in \S 2, bending waves are strongly damped at vertical
resonances where the wave frequency is nearly equal to the vertical
oscillation frequency of disk particles (equation \ref{eq:sdef}).
The epicyclic frequency, $\kappa_z$, for small-amplitude oscillations is
plotted for this model in Figure \ref{fig:freq}.  The forcing frequency
is equal to $\kappa_z$ at $R=3.43$.  This is somewhat outside the radius
where maximum disk thickening occurs, $R \simeq 2.5-3.0$.  This asymmetry
is expected for outward propagating waves, since equation (\ref{eq:sdef})
shows that significant damping will occur when the vertical frequency is
within a few times $k\sigma$ of $\omega$.

This simulation suggests that disks are heated when a perturber excites
bending waves that travel across the disk until their energy is absorbed
at the location near to the 1:1 particle resonance.

\subsection{Bowl modes}

We have attempted to reproduce the axisymmetric ``bowl mode'' predicted by
Sparke (1995) for cold, razor-thin disks inside rigid halos, but without
success.  These modes are expected to be damped when the halo can respond
(Nelson \& Tremaine 1996), but in realistic disks they are, in fact,
damped much more effectively by the disk stars.

In experiments in which we held the halo rigid, we found that a radially
warm and thick disk no longer supports this mode because its frequency,
which is set largely by the halo, is usually in resonance with the vertical
motion of the particles somewhere in the disk.  Initial bends having the
shapes of the mode predicted by Sparke decay rapidly causing localized
heating even when the halo mass is held rigid.  The decay rate was very
similar when we substituted a live halo, indicating that all the significant
damping occurred in the disk.

\subsection{Some puzzles}

We have extended the series of experiments with live halos from
\S\ref{sec:forced} to other forcing frequencies.  At most of the forcing
frequencies we used we obtained a response not only at the driven frequency
but also at a second, well-defined frequency.

When the forcing frequency was high enough for the waves to be Landau
damped by the disk particles, we generally obtained two heating peaks,
whereas only one was expected.  The result shown in Figure \ref{fig:forced}
is actually quite exceptional.  Whenever a second peak occurred, Fourier
analysis of the displacements of the disk over time always revealed
a second frequency with a resonance where the heating was observed.
The extra response frequency was neither a simple multiple nor fraction
of the driven frequency.  Moving the driving disturbance to a different
radius did not change the extra response frequency, nor did changes to
its amplitude.  In some cases, the heating caused by the extra response
exceeded that caused by the driven wave.

At still lower frequencies, for which no 1:1 resonances were present in the
disk, we again obtained two responses.  With a forcing period $\tau=80$,
for which the 1:1 resonance is right at the disk edge, we were surprised
to find that the displacement of the entire disk instantly began to behave
as the super-position of two neutral, large-scale bending waves of similar
amplitude.  The two waves have different frequencies giving rise to beats.
We have estimated the two waves to have frequencies $0.077$ and $0.068$
with negligible growth/decay rates; the higher is the forcing frequency
but the lower appears to be a natural oscillation frequency of the whole
disk-halo system.  We found that the system abruptly stops oscillating
as soon as the forcing is turned off.  At even lower forcing frequency,
$\tau=120$, the disk-halo ``mode'' seemed to damp quite quickly, leaving
the system to oscillate steadily at the driven frequency.

We have no explanation for these extra frequencies, but believe they are
not numerical artifacts since we could reproduce them with other codes.
The referee, J Goodman, suggested that they could result from non-linear
coupling to an axisymmetric density wave, but we were unable to find any
evidence for density waves with appropriate frequencies.  We consider
this strange behavior to be irrelevant to the conclusions of this paper,
since we have seen it only in the artificial circumstances of harmonic
external driving.

\section{Discussion and conclusions}

The goal of this paper has been to gain physical insight into disk heating
and thickening caused by the accretion of small satellite galaxies.
We have chosen to explore an axisymmetric system in order to isolate the
vertical heating phenomenon.  This strategy has the advantage that our
simulations are inexpensive, allowing us to sample a broader region of
parameter space while avoiding internal evolution and artificial heating
from numerical relaxation.  We believe that most of the conclusions below
also apply to the general case of off-axis satellite accretion.

We have found that the satellite loses little energy to direct heating
of disk stars, because most of the power in the satellite force is at
frequencies lower than the natural oscillation frequencies of the disk
stars. There is substantial heating at late stages, however, through the
intermediate process of exciting disk bending waves.  Once excited, these
waves eventually damp efficiently at wave-particle resonances, thereby
heating the disk non-locally.  The only significant radial heating also
occurs at the time the disk is thickened by the bending waves.  Thus,
satellites which are tidally disrupted before they are able to excite
bending waves do not thicken the disk.

Bending waves can damp by several mechanisms, including dynamical friction
from the halo, nonlinear damping at the disk edge, and internal wave-particle
resonances.  In most cases wave-particle resonance (Landau damping) is by
far the strongest damping mechanism; since this mechanism depends strongly
on the disk thickness and vertical frequencies, numerical simulations must
accurately reproduce these quantities in order to represent the behavior
of real galaxy disks.

Thus collective effects can significantly reduce the vertical energy
deposited by the satellite only through tilting the disk in an off-axis
encounter, as shown by Huang \& Carlberg (1997; see also Athanassoula 1996).
Tilting the target galaxy, reduces the vertical energy of the satellite
about the new disk plane.

We have not explored many important issues discussed by TO and others;
for example, we have not simulated the full range of satellite impacts
with arbitrary orbital orientations and satellite masses, and we have
not attempted to estimate the average heating rate.  Since off-center
encounters can excite disturbances with $m\ne0$, which also contribute
to heating the disk, we cannot address the actual extent to which disk
heating is important in real galaxies.

Many effects that we have ignored, however, such as tidal stripping of
the satellite, late star formation, and tilting the disk, tend to make
the satellite less destructive.  Thus, the heating we obtain is an upper
limit that should result from an axial encounter with a low-mass satellite.

This work was supported by NSF grants AST 93/18617 and AST 96/17088 and
NASA Theory grant NAG 5-2803 to JAS and was begun during a long visit by
JAS to CITA, whose hospitality is gratefully acknowledged.  We thank Alar
Toomre for many discussions and much advice over an extended period and
Jeremy Goodman for a thoughtful referee report.

\end{document}